\documentclass[12pt,preprint]{aastex}

\shorttitle{Formation of Giant Planets in Vortices.}
\shortauthors{Klahr and Bodenheimer}

\begin{document}


\title{Formation of Giant Planets by Concurrent Accretion of Solids and Gas inside
an Anti-Cyclonic Vortex}


\author{Hubert Klahr \altaffilmark{1}}
%\email{klahr@mpia.de}
%\email{VERSION: 2.2 Wed., Dec. 18th 2003}
%\email{STATUS: manuscript}

\affil{Max-Planck-Institut f\"ur Astronomie, Heidelberg, Germany}
\email{klahr@mpia.de}
\altaffiltext{1}{also at: UCO/Lick Observatory, University of California,
    Santa Cruz, CA 95060}
\and
\author{Peter Bodenheimer}
\affil{UCO/Lick Observatory, University of California,
    Santa Cruz, CA 95064}
\email{peter@ucolick.org}


\newcommand{\tramp}{{\sc Tramp\,\,}}
\newcommand{\okappa}{{\tilde{\kappa}}}

\begin{abstract}
We study the formation of a giant gas planet by the core--accretion
gas--capture process,  with numerical simulations, under the assumption
that the planetary core forms in the center of an anti-cyclonic vortex.
The presence of the vortex concentrates particles of centimeter to meter
size from the surrounding disk, and speeds up the core formation process.
Assuming that a planet of Jupiter mass is forming at 5 AU from the star,
the vortex enhancement results in considerably shorter formation times
than are found in standard core--accretion gas--capture simulations.
Also, formation of a gas giant is possible in a disk with mass comparable
to that of the minimum mass solar nebula. 
\end{abstract}




\section{Introduction}
Two different  models for the formation of  gas giant planets in a disk
of gas and dust are under discussion.
Both are subject to major difficulties. A gravitational  instability
 in a disk (Boss 1997) would be a fast process but probably
is not achievable, except in the outer regions of a disk (beyond 
$\approx 30$ AU; see \S 2.1), 
 because of the requirement that the cooling rate
by radiation be comparable to or shorter than the orbital time scale
(Gammie 2001; Rice et al. 2003). 
The core accretion model,  on the other hand,  explains a number of 
observed properties of the giant planets in the solar system,  but for
standard low-mass disk parameters it  requires a time scale that is
longer than the expected lifetime of
the solar nebula. In order to explain the formation of Jupiter
within 3 million years or less (Hubickyj, Bodenheimer, \& Lissauer 2005),  
one has to postulate an enhancement in the surface density of solid
particles in the disk well above (by a factor of $\sim 3$)
 that in the minimum mass solar nebula (MMSN; Weidenschilling 1977, Hayashi 1981). 
Other detailed studies of the accretion
of the core of Jupiter (Weidenschilling 1998; 
Inaba, Wetherill, \& Ikoma 2003; Thommes, Duncan, \& Levison
2003) suggest that the 
disk has to have up to 10 times the mass of the MMSN. 
Formation of the Oort cloud and the outward migration of Uranus and Neptune
imply the removal from that region of about 50 M$_\oplus$ of condensed
material, which is comparable to the core masses of all four
giant planets combined. Thus a more realistic minimum mass would be 
about twice the classical value, which, however, is still insufficient 
according to the recent simulations. 
The discussion of an acceleration mechanism is the main point of this paper. 
We wish to show that even in a MMSN it could be possible to form
Jupiter in the required time, just to show how efficient the mechanism
proposed in this paper can be. 

\subsection{Formation of vortices}
   The vortices can either form primordially by perturbations
induced by matter infalling onto the disk (Barranco \& Marcus 2000),
or result from a hydrodynamical instability, such as  the Rossby wave
instability (Lovelace et al. 1999; Li et al. 2001)
or a global baroclinic instability (Klahr \& Bodenheimer 2003a,b). 
Entropy gradients
in rotating systems can produce Rossby waves, which can eventually break into 
vortices, as is known to occur near planetary surfaces in the solar system, and
in the atmospheres of giant planets.

{Until recently anti-cyclonic vortices were thought to 
be excluded by the action of magnetohydrodynamics (MHD) and
that the magnetorotational instability (MRI) would destroy any 
large scale vortices.
Recent simulations by Fromang \& Nelson (2005) showed the opposite.
They performed three-dimensional global MHD simulations of
protoplanetary disks in a cylindrical potential, very similar to previous studies by Armitage (1998), Hawley (2001) and Steinacker \& Papaloizou (2002).
But Fromang \& Nelson (2005) measured for the first time the vorticity
distribution of the flow and found that large-scale and long-lived 
anti-cyclonic vortices formed.
These extended vorticity minima are easy to miss as 
the short-lived and small-scale vorticity fluctuations
are stronger in amplitude.
They also found that up to 75$\%$ of all
meter sized objects  (referred to here as ``boulders")
randomly placed into the disk radially outside of the vortex 
were captured in those vortices for
 a simulation which ran for a time of 200 orbits at the inner edge of the disk.
Less than half the boulders got lost and migrated off the grid,
potentially into the star.
We will again refer to this very important result for the discussion
of the capture probability of boulders by vortices.}

\subsection{Stability of vortices}

   Two-dimensional anti-cyclonic vortices can survive in protoplanetary disks
without strong turbulence,  that is, with low viscosity, for many orbits
 (Godon \& Livio 1999a,b). 
The analytic solutions for large three-dimensional  vortices (Goodman, Narayan \& Goldreich 1987)
 have recently been numerically 
tested for stability by Johansen, Andersen \& Brandenburg (2004).      
They find that the lifetime of the  3-D  vortex is inversely
proportional to an artificially imposed  background viscosity.

{ Recently, Barranco \& Marcus (2005) studied the 
stability of three-dimensional
vortices in stratified disks using an anelastic approximation. 
They find that an analytic vortex solution
to the anelastic equations 
is not hydrodynamically stable if placed in the midplane of a disk.
But these small subsonic\footnote{The anelastic approach can only handle
subsonic flows.} 
vortices are quite different from the large 
azimuthally extended vortices that show
up in Klahr \& Bodenheimer (2003a) and in Fromang \& Nelson (2005).
The vortex used by Barranco \& Marcus (2005) was placed into
the midplane,  but it decayed and reappeared in the stratified atmosphere 
away from the midplane. 
This indicates that stratification is actually
benefitial for the stability of the vortex.
In fact a vortex seems to be stable if it is flat, e.g.\ the vertical
extent is smaller than the radial and horiziontal extent. Stratification
is such an agent to keep vortices flat. Another possibility to allow for
flat vortices is to use global models beyond the limits of a shearing sheet box.
%Thus one could conclude that
%vortices cannot survive if placed in an unstratified disk.
Fromang \& Nelson (2005) showed that a three-dimensional 
vortex does survive in a global non-stratified disk.
Thus, it is possible that the instability in Barranco \& Marcus (2005)
results from the limitations of the anelastic and of the local approach, 
while global and fully compressible
hydrodynamics and MHD indicate a certain stability for three-dimensional
anticylonic vortices.}

%If protoplanetary disks have indeed a dead zone, where magnetic fields are
%not effective at generating turbulence, 
%as suggested  by Gammie (1996) and  Fromang, Terquem \& Balbus (2002),
%  the effective viscosity values will be
%lower than $\alpha = 10^{-4}$.   In this situation the vortex can survive
% for thousands of orbits 
%before it gets dissolved, even if there is no further generation of vorticity in the disk.

\subsection{Vortices in planet formation}
It has already been suggested that anti-cyclonic vortices can be beneficial 
for the planetary formation process (Adams \& Watkins 1995, Barge \& Sommeria 1995;
Tanga et al. 1996; Godon \& Livio 1999b).
In general the vortices can be regarded as preferred
 formation sites of planets because they
tend to collect solid particles from the surrounding disk.
If one invokes a three-phase model for the planetary formation process (Klahr 2003; Klahr \& Bodenheimer 2003b,c)
one can regard the formation of vortices as phase one. Phase two is characterized by
the accumulation of solids in the center of vortices and by the growth of a
planetary core,  and phase three  by the accretion of gas onto the core.
In this paper we investigate the latter two 
phases in the framework of the core accretion model (Pollack et al. 1996). 
This model allows us to quantify the
reduction in formation time of Jupiter by accelerated solid accretion into vortices,
a radial expansion of the feeding zone,  and  a relatively efficient process of   
 emptying the feeding zone. 

We do not claim  that planetesimals cannot form without
a vortex, but vortices
accelerate the planetesimal formation. For giant planets there
is a time scale problem, to form the entire planet before the
disappearance of the gas,  which can be overcome by this mechanism.
For terrestrial planets that particular time scale  problem does not
exist, and we do not deal in this paper with the numerous
other problems of terrestrial planet formation.

Section 2  gives  an overview of the current formation scenarios
under discussion and shows  how the new scenario fits into the old
framework.
In Section 3 we derive the accretion rate of solid material
as  an input parameter for the core accretion model. This model
is then analyzed, in Section 4,
through numerical simulations, as a possible way to form Jupiter rapidly. 
Section 5  summarizes and discusses  our results and
gives an outlook  for  future work.

\section{Planet Formation Models}
We introduce  the new three-phase planet formation scenario 
 because we think that the current models under discussion have
substantial physical problems. The new model is not a revolutionary
approach,  but more a merging of the fundamental ideas from the 
existing models plus some vortex theory from geophysical fluid dynamics.
Before we explain the new model we briefly describe the 
existing models and elucidate in more detail what the difficulties are.
Further discussion of these models can be found in a review by
Bodenheimer \& Lin (2002). 

\subsection{Disk Instability Model}
Boss (1997,\ 2001) proposed a model in which a
cool  protoplanetary disk, with roughly 10\% of the mass of 
the star,  fragments  as a consequence of its 
self gravity and spontaneously forms multiple 
giant planets within a few orbital periods. 
 Mayer et al. (2002) showed, in a numerical simulation with simple 
assumptions regarding cooling, that fragments can form and survive
if the Toomre $Q$ parameter is low enough, in their case about 1.4.
The fundamental problem with this model is that it needs an almost
isothermal gas to allow for the collapse and to keep the Toomre parameter
low. Gammie (2001, see also Rice et al. 2003) shows that
the cooling time $t_{\rm cool}$ has to be smaller than
an orbital period $t_{\rm orb}$.
\begin{equation} 
t_{\rm cool} < \frac{1}{2} t_{\rm orb}.
\end{equation} 
Radiative diffusion within a dust-rich disk as well as in a self-gravitating 
blob is
too inefficient to allow for such high cooling rates. Also
the recently invoked thermal convection can only slightly increase the
efficiency of the cooling (Bell et al. 1997); they show that
the maximum fraction of the energy carried vertically outward
through a disk  by convection is about 20\%. 
Convection  in itself does not result in energy loss from the
 disk; it  only transports the
heat to the surface where it still has to be radiated away.
Even if strong convection would be able 
to transport energy in the interior of the disk (Boss 2004)
the gas would still have to radiate at the surface as a black body, at a
maximum rate  proportional to $T^4$, where $T$ is the central
temperature in the disk. 
The cooling time would then be longer than an orbital period for
radii out to 33 AU:
\begin{equation} 
t_{\rm cool} \approx \frac{\Sigma c_v}{2 \sigma} T^{-3} =
 \frac{\Sigma}{10^3~{\rm g~cm}^{-2}} \left(\frac{T}{50 {\rm K}}\right)^{-3} \times 190 {\rm~yrs}.
\end{equation}
Here we used typical  surface densities $\Sigma$ and temperatures $T$ from the Boss model (Boss 2001);
$c_v$ is the specific heat and $\sigma$ is the Stefan-Boltzmann 
constant. Note that for even only slightly smaller temperatures the
cooling time increases dramatically.
We argue  on the basis of  this estimate that it is possible
to form brown dwarf companions out of an extended
circumstellar disk at large radii but not planets
at the distances of Jupiter or Saturn.
Basically one can conclude that the conditions needed for planet formation
by gravitational instability are unlikely to occur in disks. 

\subsection{Core Accretion}
In the other formation model 
(e.g.\ Pollack et al. 1996; Wuchterl, Guillot \& Lissauer 2000),  first 
a solid core of at least 5-10 $M_{\oplus}$ is  built
up via successive collisions of planetesimals.
Starting from this mass the core can accrete gas
from the disk. The  gas accretion rate is determined by 
the cooling of the protoplanet. This means that as long
as there are still planetesimals plunging into the
planetary atmosphere and releasing their potential energy
 within, the planet  contracts only slowly and  can not accrete gas efficiently.
Only after the mass of the gaseous envelope approaches that of
the solid core  can the contraction of
the envelope take place fast enough to provide a runaway 
gas accretion.  Even for massive
disks with three times the solid surface density  of the MMSN,
 it takes roughly 6 Myr to form Jupiter (Hubickyj et al. 2005), if the grain
opacity in the planetary envelope is based on an interstellar size 
distribution.
However, their calculations also show that  by
artificially stopping the accretion of solids at an early
time after the core is established,  or by reducing the grain  opacity in 
the envelope of the protoplanet  below the interstellar value, it is 
possible 
to start the runaway phase earlier. The reduction in grain opacity
is consistent with the detailed calculations of Podolak (2003). 
 Also a modest  enhancement of
solids in the nebula, above three times that in the MMSN,
  can reduce the growth time substantially. 

    This model contains another so far unsolved problem. Dust can efficiently
 coagulate to build boulders up to a size of tens of centimeters (Blum \& Wurm 
 2000).
 Sedimentation and radial drift due to gas drag and the sub-Keplerian rotation of the disk
 provide high enough relative velocities for
 efficient capture of smaller particles. At the same time the impact 
 velocities are low enough so that the surface of the boulder is not destroyed
 by sputtering and fragmentation (e.g.\ Leinhardt, Richardson, \& Quinn 2000) faster
 than it is supplied with mass. 

 But by the time the boulder reaches meter size,
it will drift with velocity up to 100 m s$^{-1}$ at 1 AU  (Weidenschilling \& Cuzzi 1993),
 or 22 m s$^{-1}$
at 7.5 AU (which is the case we consider in \S 3.2). 
 At velocities above 10  m s$^{-1}$  the boulder gets eroded
 by the many impacts of smaller grains due to the high relative velocity 
 with respect to the smaller objects. If this would not be a problem on its own,
 there is another effect. At a velocity of 100 m s$^{-1}$ the boulder
 will drift 
 into the sun faster than it can grow to a large enough size to be beyond the 
regime  of high radial drift velocities.
Thus the mechanism for forming kilometer-sized planetesimals is still an open 
question. 

 To be fair,   in general there is no strong observational evidence
that distinguishes between core accretion models and gravitational
instability models, except that the correlation of the incidence of extrasolar
planets with metallicity of the host star 
 (Santos et al. 2003; Fischer \& Valenti 2005) does favor core accretion models somewhat. 

\subsection{The Role of Vortices}
Vortices combine the previous models in the following sense: 
anti-cyclonic vortices stabilize mass concentrations
 in the protoplanetary disk, 
for example the blobs in the Boss model,  against shear.  In this case the
stabilizing effect is not
gravity but the action of Coriolis forces.
The resulting flow is in a so-called geostrophic balance, between 
pressure effects and Coriolis effects. 
The vortices are stable flow features
even without cooling,  and  they do not need self gravity to
stay bound (Goodman et al. 1987). This is the same effect that stabilizes 
hurricanes on earth or the giant red spot on Jupiter.

We suggest three reasons why vortices may be important  for planet formation.
In contrast to the sub-Keplerian gaseous 
disk,  (1) at least the centers of vortices orbit at the Keplerian rate, 
(2)  they have no vertical shear, and (3) 
they have  no radial shear and thus  will not generate (MHD)-turbulence.
The following  subsections amplify these points briefly. 

\subsubsection{Vortices move at Keplerian speed}
First one has to define the center of the vortex, which we take to 
be given by the maximum in pressure, and which we refer to as the ``eye". 
At a local pressure maximum the radial and azimuthal
pressure gradients vanish. Thus the only forces 
determining the motion of the gas at the eye of the 
vortex are gravity and centrifugal force. Thus the eye
must move on a Keplerian orbit and is not bound to the
sub-Keplerian motion of the gas outside the vortex.

For the vortex to generate a local pressure maximum
in the disk, it must exceed some minimum strength, or
amplitude, in terms of vorticity. 
 While all cyclones are low-pressure regions
and all anticyclones high-pressure regions, just
as in the earth's  atmosphere, a weak vortex will not
generate a local pressure maximum in the radially
falling pressure profile in a disk. 
However    the anticyclones  reported in the numerical simulations of 
 Klahr \& Bodenheimer (2003a)
are between 30 \% to 100 \% higher in pressure than the ambient
gas, and they have a chance
to decouple from the gas with the eye on a Keplerian orbit. 

   What are the consequences of this? Once  small solid particles  have
accumulated in the eye of the vortex and have grown to kilometer size, 
they will decouple from the gas and no longer actively
be bound to the vortex eye by the vortical gas motion. However they
also will be in Keplerian motion, along with the eye. 
Thus,   as long as there are no additional effects scattering
them out of the vortex eye,  the planetesimals may well stay in co-orbit with
the vortex for many, many orbits.

{Simulations by Klahr \& Bodenheimer (2003c) and Fromang \& Nelson (2005) have
shown that meter-sized boulders stay captured once they are inside a vortex for 
hundreds of orbits even if the overall disk flow is strongly turbulent.
Yet it is still to be shown what happens to objects large enough to 
decouple from the gas flow. They might  drift out of the vortex due
to some residual non-Keplerian drift of the vortex or even be scattered out
of the vortex by the gravitational torques of the disk gas (Fromang \& Nelson 2005).
The fate of kilometer sized and larger objects definitively needs
 further investigation.}

\subsubsection{Vortices have no vertical shear}

The vertical shear in an accretion disk
is purely an effect of the stronger sub-Keplerian motion
of the gas in the midplane than in the upper layers of the
disk, because the pressure is the highest in the midplane
and thus the radial pressure support the strongest.
In contrast,  all  the way through the vertical extent 
of the vortex eye, the gas will move at the
Keplerian frequency. Of course this vertical direction
is not the vertical direction of a cylindrical
coordinate system but is  given by the local effective gravity
in the  system co-rotating with the vortex eye, which means
that the  rotational axis of the vortex bends
slightly towards the rotational axis of the accretion disk with
increasing height above the midplane. 
In a thin accretion disk this  effect may well be unobservable.

   The effect of the non-existence of a vertical shear is very 
interesting. 
It means that solids can sediment to the midplane and
concentrate  to a density higher than the  critical value where
 self-gravitational effects become important, 
without generating a shear layer instability (Cuzzi et al. 1993).

\subsubsection{Inside a vortex there is potentially no (MHD)-turbulence}
As is known for hurricanes on earth, the eye of 
a (anti-) cyclone is quiet. There is probably 
little or no turbulence acting in the center of
a vortex, because  shear is required for the generation of,
for instance, the magnetorotational instability.
This radial shear is not present in a giant vortex.
   The vortices are also likely to be in a thermodynamical
equilibrium owing to their relatively small dimensions, so  that
a baroclinic instability is also unlikely.

    As a result of this absence of turbulence there
will be very small RMS velocities between the 
boulders. Also,
 collisions will be gentle, and the likelihood
of scattering out of the vortex is small.

It was already suggested that vortices
could be the direct precursors of planetary formation.   
The planets could form either by concentration of dust in the centers of
the vortices, as was suggested by Barge \& Sommeria (1995),
or by sufficient gas accretion onto a vortex so that it undergoes 
gravitational collapse (Adams \&  Watkins 1995).
The latter possibility seems unlikely in the light of our observation
(Klahr \& Bodenheimer 2003a) 
that the vortices are still far away from a critical Jeans mass.

   Basically we think that there are three possibilities concerning  what
happens to the solids once they are captured in the eye of
the anti-cyclone. 
\begin{enumerate}
\item{}The naive picture assumes that all 
captured solids will contribute to one single growing
core. This picture has the possibly significant problem that
the core
might actually leave the vortex  once it  grows to kilometer size
and decouples from the gas.
Even though  we stated that a strong geostrophic vortex will
orbit at the Keplerian rate, as will the kilometer-size planetesimal,
there are two sources of danger. 
First,   the vortex is a dynamical feature, and it could migrate in the
radial direction
by interaction with the ambient disk.  Second, even when the core
forms from material with basically the same angular momentum
as the vortex eye, a small variation in the specific kinetic
energy in the azimuthal direction can lead to a  slow azimuthal
drift of the core out of the eye of the vortex. This problem
might be overcome once one starts to investigate the feedback
of the core on the gas, via gravity as well as via friction.
These effects might stabilize once more the gas around the core.

\item{}In a second model one can assume that the boulders (meter size) 
that accrete into the vortex interior do  not 
accumulate in the center and  form one giant core, but that they 
form a ``core zone", enriched in solid mass but still containing some gas.
This particle layer could then eventually undergo gravitational collapse (Goldreich \& Ward 1973),
which in this case will not be prevented by vertical shear. The precise conditions
 for this instability  will have to be derived elsewhere;
in particular, the velocity dispersion of the accumulated particles
has to  be investigated in detail.
{A similar study for solids immersed in local MHD turbulence has been
performed by Johansen, Klahr \& Henning (2006),  who discuss the
possibility of gravoturbulent formation of planetesimals in small-scale
short-lived vortices.
This study should be expanded to a global simulation.}

   This picture has the benefit that all boulders might stay actually captured
by the vortex until the core forms in one single collapse. Thus even if    
 the vortex is not 100 percent precisely  Keplerian,  or if it radially migrates.
the solids would follow the vortex and not get ejected.     

\item{}In any case we do not expect the physical capturing process to be
perfect. Thus probably a smaller or larger fraction of planetesimals that
have decoupled from the gas may get ejected from   the vortex. 
If this is a minor effect, then this is a wonderful way to 
produce planetesimals in vortices and then scatter them to other
regions of the disk, where they could be used in 
the formation of other planets, that form
independently of a vortex. 

   If this scattering of the planetesimals out of the 
cyclone is more the rule than the exception, then it 
will become unlikely to form a planetary core mass inside the vortex.
Nevertheless the process would produce 100 m to 1 km planetesimals
which are difficult to form by any other means because of the effects
of gas drag.
Once  scattered out of the vortex, the planetesimals 
will stay at about the same radius. These planetesimals, whose total mass
would be 10--20 M$_\oplus$ or more, 
would thus still be in a radially confined and strongly enriched feeding-zone 
and could  accumulate to a core by collisions and gravitational focussing, as 
is assumed in the classical picture (Pollack et al.\ 1996).
 Thus the formation of a core for  a giant planet is  more likely at the
radial position of a vortex, as the formation time for the core
becomes reasonably small. Thus even if a planetary core does not
form in a vortex, the presence of the vortex may be 
 very beneficial for planet formation.
\end{enumerate}

The third idea will have to be elaborated elsewhere.
For this paper we will follow pictures one and two or
an arbitrary mixture of both. For our model it is
not important how the core forms, as long as the mass
of the solids inside the core zone starts to gravitationally act on
the surrounding gas in the eye of the vortex. Thus we simply 
assume that sooner or later the accreted solids will provide
the potential well for the giant planet formation.

We propose that the formation of planets is probably characterized by three phases,
that depend directly on each other:
\begin{itemize}
\item{Phase 1: Formation of anti-cyclonic vortices as pre-protoplanetary condensations}
\item{Phase 2: Accumulation of solids into the vortices  to form
 protoplanetary cores}
\item{Phase 3: Accretion of gas onto  the protoplanetary cores}
\end{itemize}
The following section will quantify phase 2, while
section 4 will make predictions on both phases  2 and 3.

\section{Accretion Rate of Solid Material}
The accretion rate of solid material $\dot M_s$ onto the planetary
core depends on the available mass $M_{s}$ in solid material
in the feeding zone of the planetary embryo,
the time scale for particle production $\tau_p$, the time scale 
for radial drift $\tau_d$,{  and the capture probability $q_c$. }
\begin{equation}
\dot M_s = \frac{q_c M_s}{{\rm max}\left\{\tau_d, \tau_p\right\}}
\end{equation}
Whatever time scale is longer determines
the accretion rate.

{ The capture probability $q_c$ has shown to be between 50$\%$ and 75$\%$
(Fromang \& Nelson 2005; see also Klahr \& Bodenheimer 2003c).
The high capture rate  can be explained in the following way. The vortex is not a 
completely isolated entity in the disk. Even if it stretches only over 45 to 60 degrees 
in the azimuthal direction,
the rotational profile is also changed in the rest of the disk.    The 
remaining 300 degrees of the disk at the radial  location of the anti-cyclone
can be interpreted as a large and thus weak cyclone. This cyclone 
expels particles and therefore hinders their inward passage.  
It eventually transports them towards the anti-cyclone. It can be observed in the
relevant simulations (Fromang \& Nelson 2005) that 
particles randomly distributed in the disk radially outside of
the vortex first  drift to the radial location
of the vortex and then drift azimuthally towards the center of the vortex.
In the following we simply use the symbol  $M_{s}$ to represent $ q_c M_{s} $.
The consequence is that the minimum mass disk required to form Jupiter will
be increased by $q_c^{-1}$ which is probably less than 2.} 

\subsection{Mass of Solid Material}
In the standard core accretion model only mass in
the vicinity of the core, in an annulus extending outward in both
directions  from the planet to about 
4 Hill sphere radii, 
can be swept up via gravitational attraction and accreted onto the core.
The Hill sphere  radius is 
\begin{equation}
R_H = R \left(\frac{M_p}{3 M_\ast} \right) ^ {\frac{1}{3}}
\end{equation}
where $R$ is the distance to the central star and $M_p$ and $M_\ast$ are
the masses of the planet and the central star, respectively. The value
during planet formation 
is relatively small, leading to the need of a high surface density in
solids and gas for the disk.         It also grows 
only slowly with the mass of the planetary embryo, which makes 
the initial growth from less than an earth mass a long process. 
The formation process is also held up by the slow contraction
rate of the gaseous envelope, once the core has swept up most of
the available solid material. 
Models without an anti-cyclonic vortex need a surface density in
solid material about three times that of the MMSN
to form Jupiter at 5 AU  in less than $10^7$ years, 
which  is the maximum accepted lifetime of the nebula itself.

   If there are anti-cyclonic vortices in the disk, then the
accumulation of solids is no longer driven by gravity but
by gas drag. Thus, the reservoir of solids that can be accreted
is not given by the Hill radius,
 but simply by the radial separation 
of the vortices in the solar nebula. Simulations (Klahr \& Bodenheimer 2003c, Fromang \& Nelson 2005)
have shown that mobile solids are accreted into their closest inner
vortex in a short time at a 50-75 $\%$  percent efficiency.
%In the hydro simulations (Klahr 2003; Klahr \& Bodenheimer 2003b) two vortices with a 
%one to two resonance have formed,  which corresponds to a ratio     
%of $(1:1.59)$ in  the radii of their orbits. 
For this paper we consider
the inner (Jupiter) vortex to be located at 5 AU and the outer (Saturn) vortex
at 10 AU. We adopt a reasonable solar nebula model (Bell et al. 1997) 
with a total surface density of $250~{\rm g}~{\rm cm}^{-2}$ on the average from 5 AU out to 10 AU. 
These parameters correspond to a disk with an effective viscosity
of $\alpha = 10^{-3}$ and a mean accretion rate of the disk gas onto
the star of  $\dot M = 10^{-8}$ M$_\odot$ yr$^{-1}$.
With these parameters one is close to the MMSN  
and also in good agreement with the currently accepted accretion disk theory.
For a conservative solid to gas fraction of $1:100$ this corresponds to
a surface density of  $\Sigma_d = 2.5{\rm~g~cm}^{-2}$ in solid material.
 The total solid  mass between 5 and 10 AU is then:
\begin{equation}
M_{s} = 2 \pi \Sigma_d R_{mean} dR  = 1.33 \times 10^{29} {\rm g} = 22 M_\earth.
\end{equation}
This is enough mass to form a core for Jupiter, even if the 
growth and capture processes   are not 100 percent efficient, 
and also to explain its  mean  metallicity.
{ If the capture probability is as low as 50 $\%$ (Fromang \& Nelson 2005),
then one can still compensate for  this effect by a two times larger mass
in the nebula, which is still less than the 10 times MMSN required by 
previous work (Weidenschilling 1998; 
Inaba, Wetherill, \& Ikoma 2003; Thommes, Duncan, \& Levison
2003).}

\subsection{Particle Drift}
The radial drift velocity $v_r$ of particles with friction time 
$\tau_f$ sets a lower limit to the aggregation time
of a core in the anti-cyclonic vortex:
\begin{equation}
\tau_d = dR / v_r(R_{mean}).
\end{equation}
The actual time a particle of radius $a \sim 30~{\rm cm}$ needs inside the vortex to spiral to its center is
only one orbital time ($10~{\rm yr}$) (Chavanis 2000) and thus can be neglected, as we
will see later. Chavanis (2000) also calculated the escape probability due to stochastic 
motion for the
boulders and found that  30 cm particles have no chance to escape
from the vortex in the time needed for the planet formation process.

The radial drift velocity for the relevant objects, up to meter size,  is given by 
 Weidenschilling \& Cuzzi (1993):
\begin{equation}
v_r = 2 dV \Omega \tau_f,
\end{equation}
where  $\Omega$ is the orbital frequency and 
$dV$ is the deviation of the rotational velocity of the gas from the Keplerian value,
which is an effect of the radial pressure gradient in the gas.
This deviation can be estimated to be 
\begin{equation}
dV = \frac{c_s^2}{\Omega R}.
\end{equation}
The sound speed at $R_{mean} = 7.5 AU$ in our model is 
 $c_s = 4 \times 10^4~{\rm cm/s}$, 
and the Keplerian velocity is $V_{Kep} = 1.1  \times 10^6~{\rm cm/s}$. Thus
the deviation from Keplerian velocity is 
\begin{equation}
dV = 1.5 \times 10^3~ {\rm cm/s}.
\end{equation}
The dimensionless friction time ($\tau_f \Omega$) for particles of radius $a$
 and density $\rho_d$ 
in the Epstein regime is  determined by
\begin{equation}
\tau_f \Omega = \frac{a \rho_d}{c_s \rho} \Omega \approx \frac{a \rho_d}{\Sigma},
\end{equation}
for gas of density $\rho$  and  surface density $\Sigma$.
The maximum drift velocity will occur for $\tau_f \Omega = 1$ particles with $\rho_d = 1.39~{\rm g/cm}^3$ 
at the size of $180~{\rm cm}$.
We can then define:
\begin{equation}
v_r = dV \frac{a}{90 {\rm cm}} {{\rm cm}/{\rm sec}},
\end{equation}
which holds for boulders of up to $90~{\rm cm}$ in radius.
The drift time for $dR = 2.5$ AU is a function of particle size:
\begin{equation}
\tau_d = 8 \times 10^2 \frac{90~{\rm cm}}{a}~{\rm yr}.
\end{equation}
So even for bowling-ball-sized boulders ($a = 10.8~{\rm cm}$)
the radial drift time is less than $8000~{\rm yr}$.
In Figure \ 1 we  compare the radial drift time as a function of 
particle size to the characteristic growth time.

\subsection{Particle Growth}
The weakest point of our estimate is the growth time $\tau_p$ for
$1~{\rm cm}$ to $1~{\rm m}$ sized boulders starting from 
micron sized dust, or, to be more precise,
the production rate of those objects $\dot M_p = M_{s} / \tau_p$.
Smaller objects do not decouple sufficiently  from
the gas to reach a vortex in less than the lifetime of the protoplanetary disk.
Based on simulations of particle growth (e.g.\ Weidenschilling \&  Cuzzi 1993)
 and experiments by Blum \& Wurm (2000)
we assume the following scenario: particles grow first by Brownian
motion which takes only a couple of thousand years at $7.5$ AU.
This initial growth  can occur while the vortex at 5 AU is still forming, which
also will take a few times $10^4$ years (Klahr \& Bodenheimer 2003a).
Afterwards particles sediment and grow owing to sedimentation, 
which is also a fast process. Then they drift radially inward and sweep up smaller particles
until the collision velocity exceeds the fragmentation velocity of about $10^3$ cm/s.
This part of the growth  takes the  longest time,  and thus we neglect the time
used in the previous stages.
During the radial drift, the mass gain per boulder $\dot m$ is given by
\begin{equation}
\dot m = v_r \sigma_d n_s m_s.
\end{equation}
This means the accumulation of mass $\dot m$ is proportional to the relative
velocity $v_r$ between the boulder and smaller particles, the cross section 
of the boulder $\sigma_d$, the number density $n_s$ of the smaller particles in the
path of the boulder and their mass per particle $m_s$.
This expression can be simplified and leads to an differential equation for the 
particles' mass ($m$)
as a function of the reference particle size $a_0 = 90$ cm and of the ratio between the density
of the boulder $\rho_d$ to the density of the small dust material 
$\rho_s = \Sigma_s / H = 6 \times 10^{-13}~{\rm g}~{\rm cm}^{-3}$ in the gas:
\begin{equation}
\dot m = m \left(\frac{6 dV}{4 a_0} \frac{\rho_s}{\rho_d}\right).
\end{equation}
This equation  indicates exponential growth with the typical
time $\tau_g$ at 7.5 AU of 
\begin{equation}
\tau_g = \frac{90}{1.1 \times 10^3} \frac{\rho_d}{\rho_s} {\rm sec}= 6 \times 10^3 {\rm yr}.
\end{equation}
%where we assume the original solid density in the gas as $\rho_s 5 \times 10^{-13}$
%to be enhanced by sedimentation by a factor of $q = 10.0$.
The entire growth time $\tau_e$ for particles from micron to
10 cm sized objects, e.g.\ fifteen orders of magnitude 
in mass,
should be about $\tau_e = 10^5~{\rm yr}$, as a rough
estimate with the assumed physics.
{ This gives only an upper limit to the growth time. Any settling would
decrease the growth time.}

  From Figure \ 1 we see that the drift time scale $\tau_d$ is shorter than the
growth time scale $\tau_g$ for particles larger than $10$ cm.
This means after they reach a size of $10$ cm they will drift 
into the next vortex  before they  can  grow by an order of magnitude in size.
Thus, the boulders grow on the average in $\tau_e = 10^5$ yrs to the critical size, and then
they become accreted within $4 \times 10^3~{\rm yr}$ into the center of the vortex to form the core.

  It is more complicated to estimate the production rate $\sim \tau_p$ of $10$ cm solids.
Theoretically, all boulders could grow simultaneously and reach the critical size at the
same time. Then the production rate would be infinite and the drift time scale the
only limiting factor.
On the other hand there is a spread in the particle size distribution (see Weidenschilling \& Cuzzi 1993). This translates into a spread in the time that particles need to grow to the critical size
of maybe $\pm \frac{1}{2} \tau_e$.
Thus, we can conservatively estimate an
accretion rate of solids of
$\dot M_s = 2.2 \times 10^{-4} M_\earth / {\rm yr}$.
 In  case our estimate is still wrong by one order of magnitude
we also consider a model of $\dot M_s = 2.2 \times 10^{-5} M_\earth / {\rm yr}$ and compare the results.

It is interesting to note that the drift time will be smaller than
the growth time for particles of about $a = 10~{\rm cm}$ (see Fig.\ 1),
which means that no significant number of larger particles can probably form 
outside the vortices and thus, the critical velocities for
fragmentation of particles which occurs for $a = 90~{\rm cm}$ sized 
boulders will not be reached. This fragmentation was so far considered to be 
the end of planetesimal growth (e.g.\ Leinhardt, Richardson, \& Quinn 2000), and the formation
of larger objects an unsolved problem. We conclude that vortices are
not only helpful to form gas giants but also to build up planetesimals,
which will be either scattered out of the vortex later on or set free
after the vortex has dispersed.

\section{Time Scales for Gas Accretion}

The assumed mechanism for planet formation is the
core accretion -- gas capture process. The vortex is assumed
to have been formed at 5 AU from the central object.  Particles
in the 10 cm size range  migrate inward in the disk
as a result of gas drag and accumulate in the vortex
where they quickly spiral toward its center. The 
vortex is assumed to last long enough so that all the
solid particles originally  between 5 and 10 AU are
captured by it.

The procedure for numerically calculating the formation of
a planet is described by Pollack et al. (1996).  There
are three main elements of the calculation: 1) the determination
of $\dot M_{\rm s}$ by accretion of planetesimals, 2)
the calculation of the interaction of accreting planetesimals
with the gaseous envelope through which they fall, to determine
where they dissolve in the envelope or whether they reach the
core, and 3) the calculation of the evolution of the
envelope. Steps 1 and 2 are simplified in the current
calculations. 

Although Pollack et al. (1996) made a detailed calculation
of the variable accretion rate of planetesimals onto a
planetary core, in this case we simply assume that small
particles are accreted at a constant rate by the vortex,
and that they quickly fall to the center to join the 
planetary core. The accretion rate stays constant until all
the available solid material has been accreted by the vortex,
after which time $\dot M_{\rm s}$ is set to zero. This
last assumption also differs from those of Pollack et al.
(1996) in that they assume that solid accretion continues
even during the gas accretion phase. In our picture there are
no further solids available to accrete. The core is set to a 
constant density of 3.2 g cm$^{-3}$.

\begin{table}%
%\footnotesize
\begin{tabular}{l|l|r|r|r|l|l|l|l}
 Case & $\dot M_{\rm s} (M_\odot yr^{-1})$ &$  M_{\rm core,final} (M_\earth)$ &          Opacity & Formation Time (yrs)\\
\hline
%\startdata
{I}   &$2.2 \times 10^{-4}$ & 22 & high &$0.3 \times 10^{6}$\\
{II}  &$2.2 \times 10^{-5}$ & 15 & high &$1.3 \times 10^{6}$\\
{III} &$2.2 \times 10^{-5}$ & 15 & low  &$0.8 \times 10^{6}$
 %\enddata
\end{tabular}
\caption{\label{tab1} Parameters chosen in the different
models and resulting formation time for Jupiter. }
\end{table}
\clearpage

The planetesimal interaction with the gaseous envelope is not
calculated. Planetesimals are assumed to fall through the envelope 
and hit the core, or, if they become dissolved in the envelope, 
to settle to the core. These assumptions do not give results 
significantly different from those of Pollack et al. (1996).  
Thus the energy liberated at the boundary of the core by planetesimals
falling onto it is approximately
\begin{equation}
L_{\rm acc} ={{ G M_{\rm core} \dot M_{\rm s}} \over {R_{\rm core}}}
\end{equation}

The calculation of the envelope is carried out through an
implicit numerical solution of the spherically symmetric stellar
structure equations as given by Bodenheimer \& Pollack (1986). The 
energy sources are planetesimal  accretion and gravitational contraction
of the gas. The outer edge of the planet at a radius $R_{\rm p}$
 is assumed to be the smaller of
the Hill sphere radius (eq. 4) or the Bondi accretion radius
\begin{equation}
R_b \approx {{G M_{\rm p}} \over {c^2}}
\end{equation}
where $c$ is the sound speed in the disk just outside the planet. 
The surface boundary conditions at   $R_{\rm p}$  are taken to be
a constant density and temperature taken from a standard disk
model (Bell et al.\ 1997). In the cases presented here the outer
temperature $T_{\rm neb} = 150$ K, and the outer density 
$\rho_{\rm neb}= 5 \times 10^{-11}$  g cm$^{-3}$. The mass accretion
rate for the envelope $\dot M_{\rm gas}$  is determined by the requirement that 
$R_{\rm p} = {\rm min}(R_H, R_b)$. 
These radii expand as the planet gains mass, and especially at later
stages the planet tends to contract, so  mass is added until
the above requirement is satisfied.

  The equation of state of the
envelope, which is assumed to have solar composition, is provided
by the tables of  Saumon, Chabrier, \& Van Horn (1995).   The opacities
are provided by grains at low temperatures, and by molecules and other
sources at temperatures above about 1800 K, where the most refractory
grains evaporate. The highest temperatures that are encountered are in
the range 20,000 K. The  Rosseland mean
 opacities in the range 800--20,000 K are
provided by the table of Alexander \& Ferguson (1994), while those
in the lower temperature range are obtained from Pollack, McKay, 
\& Christofferson (1985). These opacities in the grain regime are
based on the assumption that they have an interstellar size distribution. 
In the absence of detailed calculations, we assume that the grains
in the vortex in the gas outside the planetary core have those
properties. However we do one calculation in which the opacity in the
grain regime (only) is reduced by a factor of 100. This assumption
is justified  by the work of Podolak (2003), who did a calculation of
grain coagulation, settling, and evaporation in the envelope of a
protoplanet, redid the opacity calculation on the basis of the
modified size distribution that he found, and concluded that the
opacities are roughly two or more orders of magnitude reduced from
the interstellar values. 

The initial condition for the simulation of the evolution of the
envelope is taken to be a core mass of 1 M$_\oplus$. The hydrostatic
solution for the value of the envelope mass at that point gives
a value of $1 \times 10^{-4}$ M$_\oplus$. During the course of the
calculation the envelope mass remains small until the core mass
has reached its limiting value, after which time the envelope mass
grows at an accelerating rate. When the envelope mass equals the
core mass, the rapid gas accretion phase begins,  during which the
total mass builds up to the final planet mass, as determined by
disk clearing or gap formation, on a time scale of only $10^4$ yr.
Since the main point of the calculation is to determine the
formation time of the planet, we stop the calculation when the
envelope mass is slightly larger than the core mass. We note that
a standard core accretion process, without the presence of the
vortex, in a disk at 5 AU with a solid surface density of 2.5
g  cm$^{-2}$, would have a formation time much longer than the
typical half-life of a disk, 3 Myr (Haisch et al. 2001). 

The parameters of the three cases are listed in Table I.  Case I
is the standard case discussed above. In Case II the assumed
accretion rate is decreased by a factor of 10, and it is assumed
that the final core mass is only 15 M$_\oplus$. Both of these
changes tend to increase the formation time, so these parameters
give a rather unfavorable case for planet formation. Case III is
identical to Case II, except the grain opacities are reduced
by a factor of 100. This change tends to speed up planet
formation during the phase of relatively slow gas accretion
(Pollack et al. 1996).

As discussed above the evolution separates fairly neatly into 
a core accretion phase and an envelope accretion phase. 
The masses of the core and envelope, as well as the luminosity
radiated at the planet's surface, are  plotted as a function of 
time for Case I in Figure  \ref{fig1.ref}.  During the first
20000 yr the value of $\dot M_s$ is gradually increased from an initial
value of $1.4 \times 10^{-5}$ M$_\oplus$ yr$^{-1}$ to its final 
constant value of $2.2 \times 10^{-4}$ M$_\oplus$ yr$^{-1}$, which it
reaches when the core mass is 2  M$_\oplus$. The luminosity at this
point is $10^{-5}  $ L$_\odot$. Thereafter 
the core mass grows linearly with time, the envelope mass is negligibly
small, and the luminosity grows as $M_{\rm core}^{2/3}$. 
Starting at about $10^5$ yr, $\dot M_s$ is smoothly reduced to zero. 
The envelope mass at this time is about 0.2  M$_\oplus$. As the energy
source arising from accretion of planetesimals is cut off, the
luminosity rapidly falls, since the only remaining source is the
gravitational contraction of the low-mass envelope. The luminosity 
levels off at just above  $10^{-6}  $ L$_\odot$.
As the envelope contracts rapidly to supply the radiated luminosity,
its mass increases according to the requirement that the outer
edge of the planet be at $R_{\rm p}$.          As the envelope mass
increases, the luminosity increases as well. At about $3 \times 10^5$
yr the core mass and envelope mass are the same, the so-called
crossover mass. Thereafter the envelope mass increases extremely
rapidly until gas accretion is terminated by gap opening or by
the dissipation of the disk. This later stage is not followed
because the total formation time will be only slightly longer than
the time to crossover. Essentially planet formation is complete
at  $3 \times 10^5$ yr, a factor of 20 faster than the standard
core accretion model would need with a considerably higher (factor 4)
solid surface density and with the same opacity. 

A plot of the same quantities is shown in Figure  \ref{fig2.ref} 
for Case II. The core accretion rate is a factor 10 slower than in
Case I and the accretion time for the core is $ 6 \times 10^5$ yr, 
less than a factor of 10 longer than in Case I because the final core
mass is only 14.7   M$_\oplus$ rather than 22 M$_\oplus$. 
At  a core mass  of 2  M$_\oplus$, the luminosity is  $10^{-6}  $ L$_\odot$, 
a factor of 10 lower than that in Case I at the same core mass.
Again the luminosity drops by about an order of magnitude just after the
core accretion is cut off, then begins to climb again as the envelope mass
approaches that of the core. The luminosity during most of the envelope
accretion phase is about a factor 5 less than that in Case 1, mainly
as a result of the smaller core mass, as explained in Pollack et al. (1996).
Thus the contraction rate, and the accretion rate of the gas, are slower
by a similar factor. The gas accretion phase lasts $7 \times 10^5$ yr, 
compared to  $1.8 \times 10^5$ yr in Case I. The overall evolution time
is $1.3 \times 10^6$ yr, a factor of 4.3 longer than that in Case I, 
but still shorter than typical disk lifetimes. 

 The only difference between Case II and  Case III (Figure \ref{fig3.ref})
is the reduction of the opacity in the envelope. The core accretion phase is
unaffected, so the time scale and luminosity are the same as in Case II.
The reduction in the envelope opacity results in an increase in the
average envelope luminosity by about a factor of 3 in comparison with 
Case II. The envelope accretion phase lasts $2.2 \times 10^5$ yr,
and the total formation time is $8.2 \times 10^5$ yr.
Note that if the same opacity reduction had been applied to Case I,
which has the standard core accretion rate, the overall evolution time
would have been  only $2 \times 10^5$ yr, which is the most likely estimate
of the planetary formation time in a vortex, since it incorporates the
most reasonable values of parameters. 

\section{Conclusion}
We present a new formation model for gas giants.
The general idea is that a giant vortex can accelerate 
the core formation considerably, even in a low-mass disk.
The envelope accretion phase is speeded up also, because once
the core has accreted all available solid material, the only energy
source available for the gaseous envelope is its own contraction.
The main reason for the long formation times for Jupiter  in the earlier
models of Pollack et al. (1996) was the additional contribution
of planetesimal accretion to the energy supply  of the envelope, 
during the first part of the gas accretion phase. 
We determine approximately,   for the first time, the resulting
time scales of such a scenario for the case of Jupiter's formation in a MMSN,
{ (or twice MMSN, if the capture probability is as low as 50$\%$)}. 
If  a vortex  had been 
responsible for the formation of Jupiter, the formation time 
would fall in  the range $2 \times 10^5$  -- $1.3 \times 10^6$ yr.

   The major drawback of our model is that there are no direct  observations
of giant vortices in protoplanetary disks so far. But observations 
are planned to look for giant vortices with future instruments such as 
ALMA (Wolf \& Klahr 2002). 
A vortex can be observed with ALMA because
it is usually a region of higher gas surface density and thus also more
dust is available to emit radiation. 
The boulders that
become trapped in the vortices are not observable, but if they collide
then some micron sized debris could enhance the dust to gas ratio and 
increase the opacity in the vortex region. But again, in Wolf \& Klahr (2002)
we show that the vortex can be
observed even without accumulating more dust.

%It has been suggested that the signature of a vortex
%has already been seen in the object KH 15D, which shows large periodic
%photometric variations (Kearns \& Herbst 1998; Hamilton et al. 2001; 
%Barge \& Viton 2003).
This paper concentrates on the second and third  of the phases
outlined in \S 2.3. It 
still needs to be proved that (1) conditions necessary for vortex
formation actually commonly occur in disks, and (2) that a vortex actually
survives long enough so that a planetary core of 10--20 M$_\oplus$ can form 
in it. Initial studies by Li et al. (2001) and Klahr \& Bodenheimer (2003a,b)
and most recently for the case of MHD by Fromang \& Nelson (2005)
indicate that vortices can form, and work by Adams \& Watkins (1995) 
and Johansen et al.\ (2004) shows that vortices
can survive,  at least  up to several $10^4$ yr at 5 AU (Godon
\& Livio 1999a,b). 
A linear stability analysis shows that vorticity can be generated
from entropy gradients in the disk (Klahr 2004; Johnson \& Gammie 2005),
which is  a necessary condition to form large scale vortices.
Johansen \& Klahr (2005) showed that in local MHD simulations
one can find anti-cyclonic vortices to form and concentrate
particles, but those small vortices which are
part of the turbulent flow, do only survive a few orbits, before they 
decay.
{ Even more convincing are the global simulations by Fromang \& Nelson (2005)
which find the formation of long-lived anti-cyclonic vortices
in MHD turbulence. Thus, it seems that vortex formation
is a generic feature to any kind of turbulence in 
accretion disks.}

Further work is required to show how robust the vortex production process is.
Assuming that the above conditions are satisfied, the  main
 benefits of the vortex-core planet formation model are:
\begin{itemize}
\item{No need for a solar nebula much more massive than minimum mass.} 
\item{Diminished loss of boulders as a result of drift into the central object.} 
\item{No fragmentation of boulders as a result of   high impact velocities.}
\item{Gentle aggregation of a core in the quiet eye of the vortex, which need not be
self-gravitating.}
\item{A formation time far less than the lifetime of the nebula.}
\end{itemize}

{ We conclude that  this model is able to solve outstanding problems in the theory 
of planet formation, and that further work on the difficult problem of vortex
generation, through  MHD                     
simulations  including radiation transport,  transport and feedback effects
 of the
solid boulders, and self gravity in three space dimensions, is warranted.}

\acknowledgments
We want to thank our referee Stuart Weidenschilling
for his constructive criticism, and 
Doug Lin and Norm Murray for fruitful discussions.
This research
has been supported in part by the NSF through grant AST-9987417,
by NASA through grants NAG5-4610,  NAG5-9661,  and NAG5-13285, and by a
special NASA astrophysics theory program which supports a joint
Center for Star Formation Studies at NASA-Ames Research Center, UC
Berkeley, and UC Santa Cruz. The 
work was  supported in part by the European Community's Human Potential
Programme under contract HPRN-CT-2002-00308, PLANETS.

\clearpage
\clearpage
\begin{figure}
\plotone{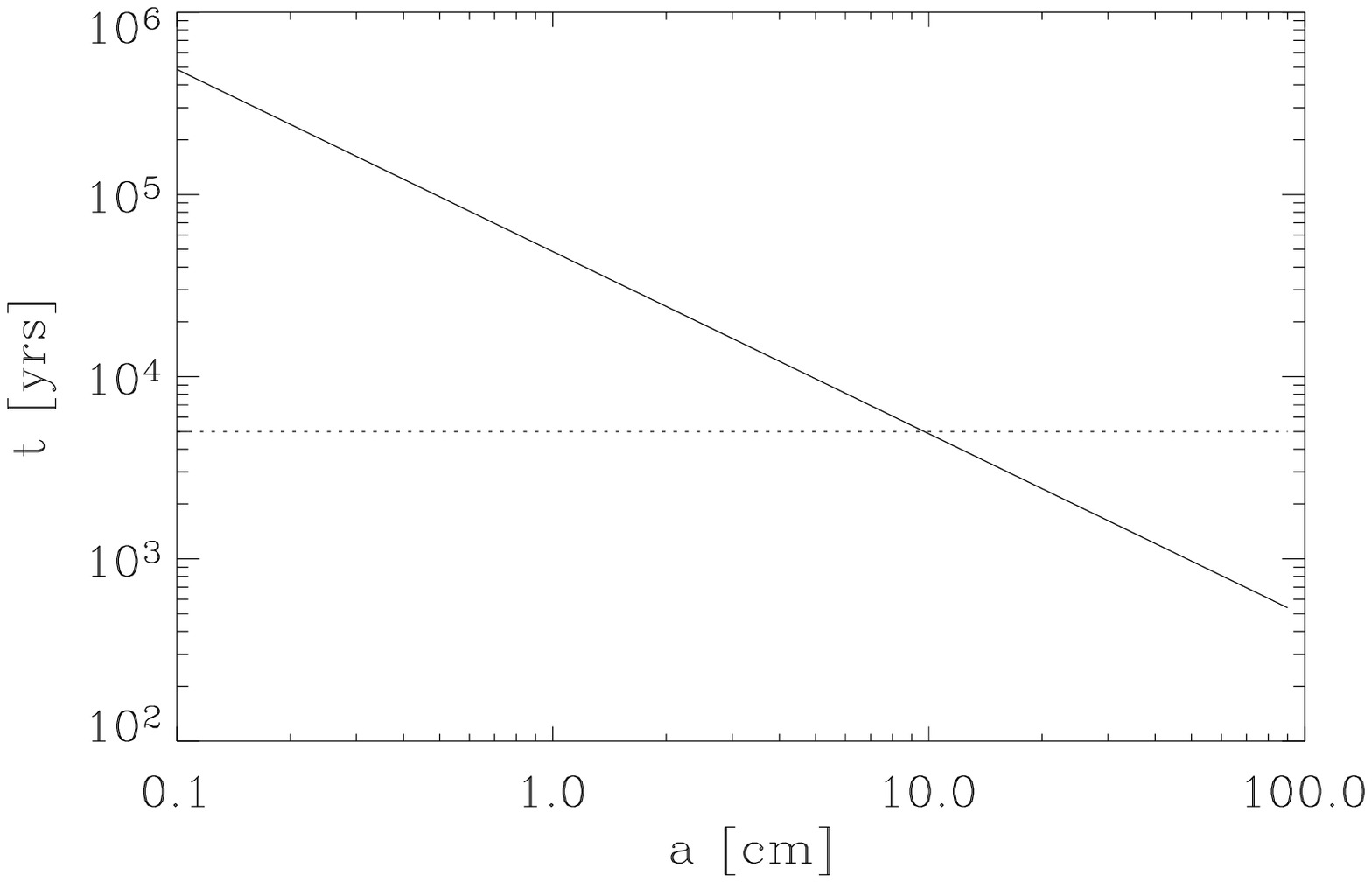}
% v1.ps
\caption{\label{fig0.ref}A comparison between drift time ({\it solid line})
 and growth time  ({\it dotted line})
for solids  as a function of size. The values are calculated using
the equations from this paper for a location of $7.5$ AU in a minimum mass solar nebula.} 
\end{figure}
\clearpage
\begin{figure}
\plotone{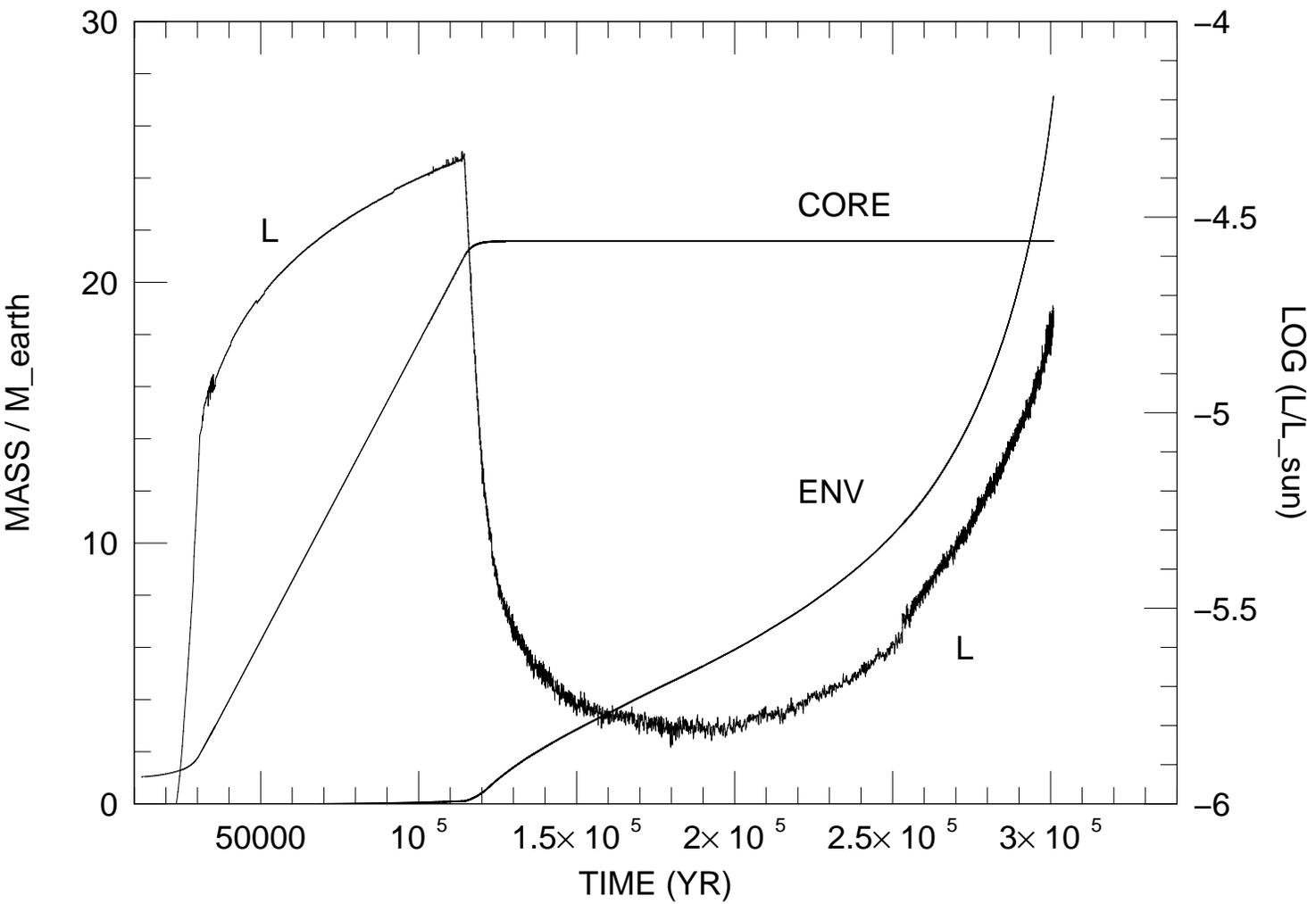}
% v2.ps
\caption{\label{fig1.ref}Case  I: Evolution of the planet's luminosity ($L$), core mass
(CORE),  and envelope mass (ENV) as a function of time.
} 
\end{figure}
\clearpage
\begin{figure}
\plotone{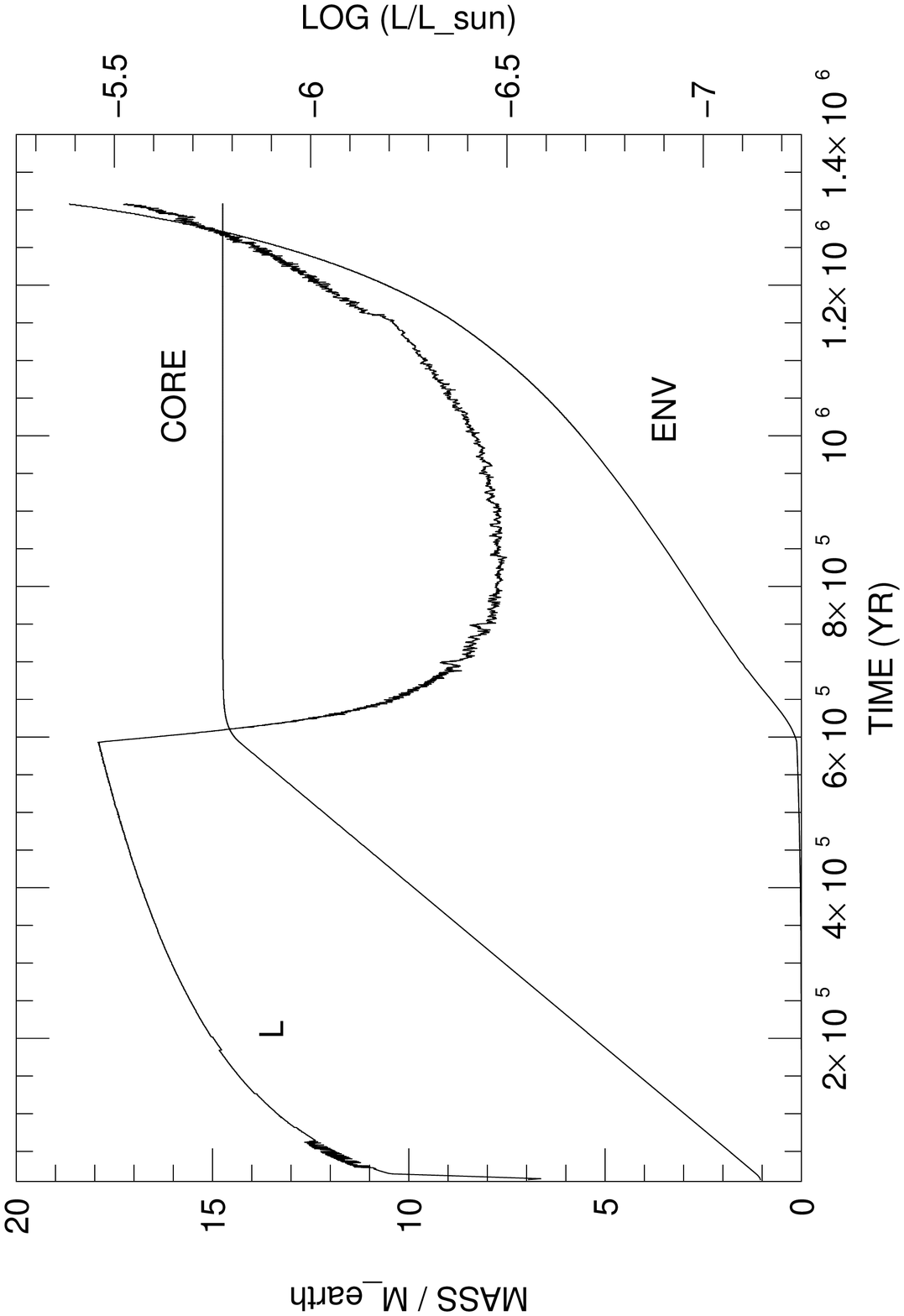}
% v3.ps
\caption{\label{fig2.ref}Case  II: Evolution of  the planet's luminosity, core mass,
 and envelope mass as a function of time. Labels have the same meaning as in Fig. \ref{fig1.ref}.
} 
\end{figure}
\clearpage
\begin{figure}
\plotone{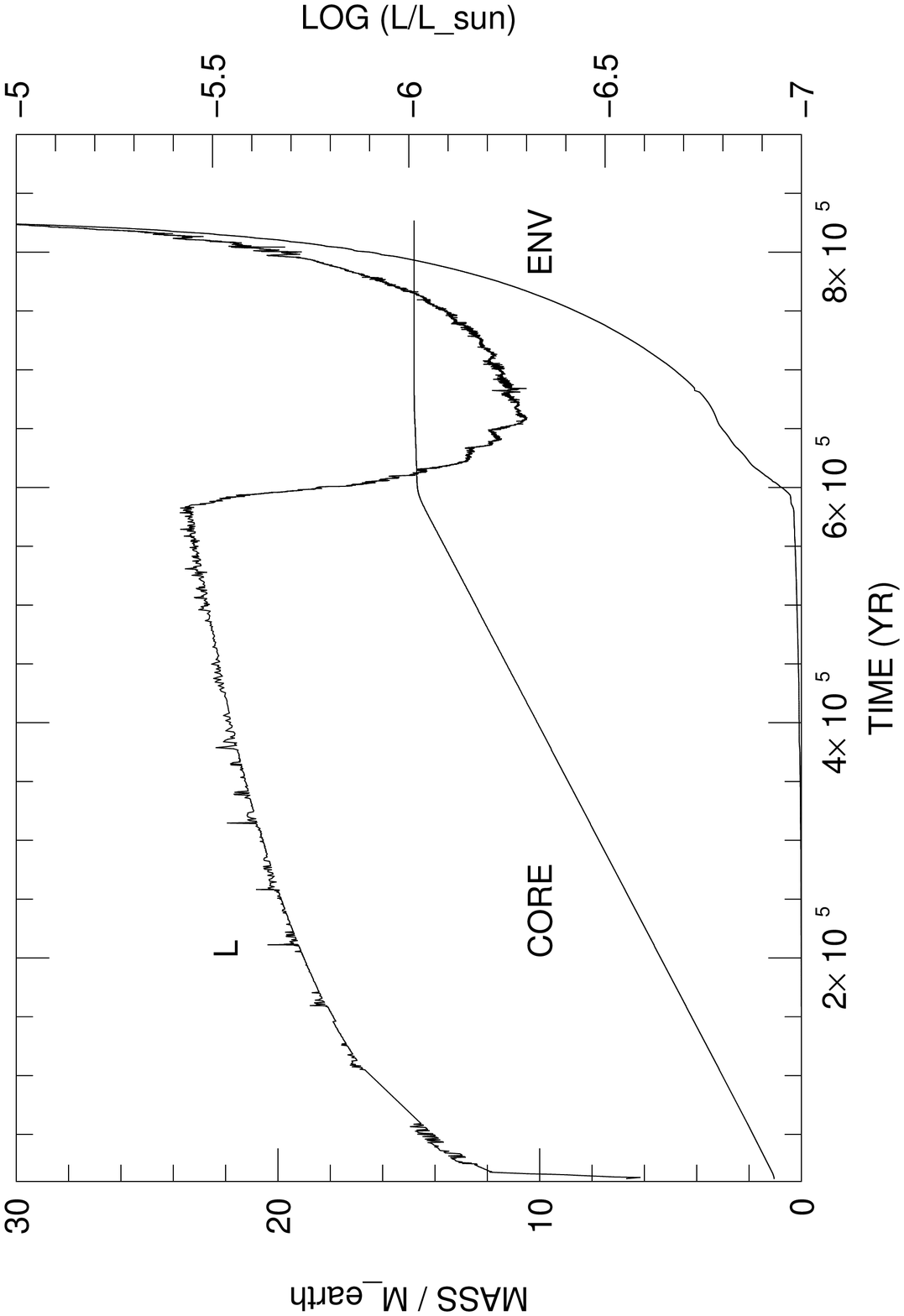}
% v4.ps
\caption{\label{fig3.ref}Case  III: Evolution of the planet's luminosity, core mass,
 and envelope mass as a function of time.  Labels have the same meaning as in Fig. \ref{fig1.ref}.
} 
\end{figure}
\clearpage

\end{document}